# Censor-Aware Semi-Supervised Survival Time Prediction in Lung Cancer Using Clinical and Radiomics Features

Arman Groji[1], Ali Fathi Jouzdani[1], Nima Sanati[1], Ren Yuan[2,3], Arman Rahmim[3,4], Mohammad R. Salmanpour[3,4,5*]

[1] NAIRG, Department of Neuroscience, Hamadan University of Medical Sciences, Hamadan, Iran
[2] BC Cancer, Vancouver Center, Vancouver, BC, Canada
[3] Department of Radiology, University of British Columbia, Vancouver, BC, Canada
[4] Department of Integrative Oncology, BC Cancer Research Institute, Vancouver, BC, Canada
[5] Technological Virtual Collaboration (TECVICO CORP.), Vancouver, BC, Canada

**(*) Corresponding Author:**
Mohammad R. Salmanpour, PhD
Department of Integrative Oncology, BC Cancer Research Institute, Vancouver, BC V5Z 1L3, Canada, Tel: 604-675-8262. Email: msalman@bccrc.ca

## Abstract

**Objectives:** Lung cancer poses a significant global health challenge, necessitating improved prognostic methods for personalized treatment. This study introduces a censor-aware semi-supervised learning (SSL) framework that integrates clinical and imaging data, addressing biases in traditional models handling censored data.
**Methods:** We analyzed clinical, PET, and CT data from 199 lung cancer patients from public and local data repositories, focusing on overall survival (OS) time as the primary outcome. Handcrafted (HRF) and Deep Radiomics features (DRF) were extracted after preprocessing using ViSERA software and were combined with clinical features (CF). Feature dimensions were optimized using Principal Component Analysis (PCA), followed by the application of supervised learning (SL) and SSL. SSL incorporated pseudo-labeling of censored data to improve performance. Seven regressors and three hazard ratio survival analysis (HRSA) algorithms were optimized using five-fold cross-validation, grid search, and external test bootstrapping.
**Results:** For PET HRFs, SSL reduced the mean absolute error (MAE) by 26.5%, achieving 1.55±0.25 years with PCA+decision tree regression, compared to SL's 2.11±0.46 years with PCA+KNNR ($p<0.05$). Combining HRFs (CT_HRF) and DRFs from CT images using SSL+PCA+KNNR achieved an MAE of 2.08±0.45 years, outperforming SL's 2.26±0.51 years by 7.96% ($p<0.05$). In HRSA, CT_HRF applied to PCA+Component Wise Gradient Boosting Survival Analysis achieved an external c-index of 0.65±0.02, effectively differentiating high- and low-risk groups.
**Conclusions:** We demonstrated that the SSL strategy significantly outperforms SL across PET, CT, and CF. As such, censor-aware SSL applied to HRFs from PET images significantly improved survival prediction performance by 26.5% compared to the SL approach.

## 1. Introduction

Lung cancer accounted for nearly 2 million cases and deaths globally by 2020, and this number is projected to rise to 3.8 million by 2050 [1]. To address this growing burden, the development of new overall survival (OS) predictive models is crucial for improving clinical decision-making, which can enhance survival rates and optimize personalized care [2]. These models also facilitate the stratification of patients into risk groups for tailored treatments, supporting the creation of targeted treatment strategies to improve patient outcomes [3, 4, 5, 6, 7, 8]. The field of OS prediction is characterized by a diverse range of methodologies, each tailored to specific clinical scenarios. Classification methods, for instance, facilitate straightforward risk stratification, allowing clinicians to categorize patients based on survival probabilities. Zheng et al. [9] demonstrated this approach with a hybrid model that integrates clinical and imaging data for predicting 2-year OS in Non-Small Cell Lung Cancer (NSCLC) patients. In the survival analysis approach, statistical methods, such as the Cox proportional hazards (CoxPH) model and machine learning approaches like random survival forests (RSF), primarily focus on modeling the hazard rate [10].

While these methods provide valuable insights into the relative risk of events, the hazard rate itself can be challenging to interpret directly in terms of actual survival time [11, 12]. Regression algorithms, on the other hand, aim to predict the exact timing of events [13]. These approaches handle censored data (instances where the death has not occurred or is unobservable within the study period) through various techniques. For instance, censored data points may be excluded [14], their influence diminished via weighting techniques [15], or addressed using Censored Quantile

Regression or parametric models like Accelerated Failure Time (AFT), which explicitly account for censored observations while estimating survival time. However, these methods often face challenges, particularly when dealing with limited data and high proportions of right-censored observations. Such challenges can lead to biased models that underestimate survival probabilities or produce skewed hazard rate estimates.

To mitigate these issues, some studies used censor-aware approaches [16, 17]. These models employ Semi-supervised learning (SSL) strategies to leverage unlabeled data and enhance prediction accuracy. The SSL techniques, in particular, show promise in addressing challenges such as missing and right-censored data, which are common in clinical research due to complexities in data collection and patient follow-ups [18, 19, 20, 21]. The SSL methods have demonstrated the potential to not only improve prediction accuracy but also save time and resources, especially in applications such as lung cancer outcome prediction [22].

Multi-omics data, including radiomics, genomics, and clinical information, play a crucial role in predicting lung cancer prognosis, such as OS, metastasis, and treatment response [23, 24]. Clinical Features (CF) like age, cancer stage, and type of surgery are commonly used for survival prediction [25, 26, 27, 28, 29, 30]. Radiomics, blending medical imaging and data science, extracts high-dimensional data from images to reflect the tumor microenvironment [31]. Features can be derived either through Handcrafted Radiomic Features (HRF) from standardized software or Deep Radiomic Features (DRF) via deep learning methods [32]. DRFs improve predictions by identifying complex patterns within imaging data and have enhanced outcome predictions in lung and head and neck cancers when combined with HRFs [33, 34, 35, 36]. However, challenges remain, including dataset quality, methodological standardization, and the "black-box" nature of DRFs. Positron Emission Tomography (PET) and Computed Tomography (CT) scans are commonly used in lung cancer diagnosis, with PET showing metabolic activity and CT providing anatomical details [37, 38].

In this study, we aim to predict OS time using HRFs and DRFs extracted from PET and CT images, along with CF using seven regression algorithms (RA). To address the limitations of simple regression approaches, which often rely heavily on uncensored data and can lead to biased results due to right-censored observations, we explored a more recent methodology of censor-aware semi-supervised regression that incorporates both censored and uncensored data. Furthermore, we assessed three ML survival analysis methods for OS prediction. These methods were shown to act more accurately compared to traditional methods for survival analysis, like the Cox proportional hazards model, especially when dealing with high-dimensional data [39].

## 2. Methods and Materials

### I. Patient demographics and clinical data

We utilized clinical data, as well as PET and CT images from a total of 199 patients, sourced from the BC Cancer database (n=166) from 2006 to 2016 and Cancer Imaging Archive (TCIA) Radiogenomics [40] (n=33) from 2008 to 2012. A detailed description of the 18 CFs utilized in our research can be found in the Supplemental Table S1. The OS time, ranging from 0.1 to 6.6 years, was used as an outcome measure and was determined by the interval between the initial diagnosis and the date of death for each patient. Among these patients, 99 had OS (6 from TCIA and 93 from BC cancer), while 100 (27 from TCIA and 73 from BC cancer) did not pass away during the study period and were included in the SSL and hazard ratio survival analysis (HRSA). Below, we present the demographic in Table 1.

**Table 1.** Demographic and Clinicopathologic Features of Patients

| **Characteristics** | | **Tumor Features** | |
|---|---|---|---|
| **Source (BC Cancer / TCIA)** | 166 (84%) / 33 (16%) | **Histology** | |
| **Sex (Male / Female)** | 100 (51%) / 99 (49%) | **Non-small cell carcinoma** | 24 (12%) |
| **Age (Mean ± SD)** | Range: 24-87 (68 ±10.11) | **Squamous cell carcinoma** | 46 (23%) |
| **Ethnicity** | | **Adenocarcinoma** | 109 (55%) |
| **Asian** | 26 (13%) | **Acinar cell carcinoma** | 3 (1.5%) |
| **Unknown** | 173 (87%) | **Neuroendocrine** | 3 (1.5%) |
| **Smoking / Not smoking** | 166 / 33 (Average of 36.4 smoking pack year) | **Combined small cell carcinoma** | 1 (0.5%) |

| Interventions | | Large cell carcinoma | 1 (0.5%) |
| --- | --- | --- | --- |
| | | Adenosquamous | 3 (1.5%) |
| Surgery Category | | Unknown | 9 (4.5%) |
| Lobectomy | 87 (43%) | Stage at diagnose | |
| Segmentectomy | 13 (7%) | IA | 32 (16%) |
| Pneumonectomy | 4 (2%) | IIA | 54 (27%) |
| Unknown | 95 (48%) | IIB | 32 (16%) |
| Chemotherapy | | IIIA | 3 (2%) |
| Yes | 59 (30%) | Unknown | 20 (10%) |
| No | 140 (70%) | Metastasis | |
| Radiation Therapy | | Yes | 64 (32%) |
| Yes | 74 (37%) | No | 102 (51%) |
| No | 31 (16%) | Unknown | 94 (47%) |

## II. Study procedure

In the pre-processing step, PET images were first registered to CT ( Figure 1, part i), then Standardized Uptake Value (SUV) correction (ii) was applied to PET images, and CT images were clipped (iii), and all images were finally normalized by min/max technique (iv) [41, 42]. These are elaborated in Supplemental section 1.1. Region Of Interests (ROIs) which represented the tumor volume were outlined by experienced medical experts to extract imaging features (v). Subsequently, two different frameworks, such as HRF and DRF, were employed for quantitative analysis. 215 HRFs were obtained from the segmented cancer regions in both PET and CT images using the standardized PySERA module in ViSERA software [32]. Moreover, 1024 DRFs were extracted from both segmented images through a 3D-AutoEncoder embedded within the ViSERA software (vi). Details of the architecture can be found in Supplemental Section 1.2. Furthermore, the CF set (vii) was combined with HRF and DRF sets to generate 13 new datasets. After new dataset generations, the features in all datasets were normalized using the min-max function, and missing values were imputed by average values (viii) and Principle Component Analysis (PCA) reduced the feature size to 10 components, retaining over 90% variance and minimizing overfitting across datasets. Subsequently, seven RAs including AdaBoost Regressor (ABR) [43], Random Forest Regressor (RFR) [44], K Nearest Neighbor Regressor (KNNR) [45], Decision Tree Regressor (DTR) [46], Linear Regression (LR) [47], Multi-Layer Perceptron Regressor (MLPR) [48], and Support Vector Machine (SVR) [49] were employed in the SL approach to predict OS with grid search optimization (ix).

As KNNR demonstrated the best performance in the SL approach, it was selected for pseudo-labeling (x) the unlabeled data. For the SSL strategy, 100 samples without events were pseudo-labeled, and then seven above-mentioned RAs were employed to train on the combined dataset, consisting of all labeled training samples and the pseudo-labeled data, using five-fold cross-validation and grid search optimization. The efficacy of the SSL method was then compared to the SL strategy, which used only the 104 labeled cases applied to KNNR as the best regressor. The MAE metric was used to compare the performance of the algorithms. Moreover, principal components derived from our datasets were applied to Fast Survival Support Vector Machine (FSSVM) [50], Component Wise Gradient Boosting Survival Analysis (CWGBSA) [51], and RSF [52] (xi), and survival risk assessments were evaluated using the C-index and Log-rank p-value on five-fold cross-validation. Kaplan-Meier survival analysis [53] was conducted to evaluate the OS differences between patients classified as high-risk or low-risk by the model, with the Log-Rank test [54] determining the significance of these differences. Moreover, we employed a mean function to categorize patients into high and low-risk groups based on their death times; patients with death times above the average were deemed low-risk, while those below the average were considered high-risk. SL, SSL, and survival risk assessments were externally validated using 100 bootstrap samples on the labeled portion of the external dataset, comprising 33 patients from TCIA (xii). A detailed description of the pseudo-labeling strategy can be found in Supplemental section 1.6, while information on PCA and the RAs, and HRSAs is provided in Supplemental sections 1.3, 1.4, and 1.5, respectively.

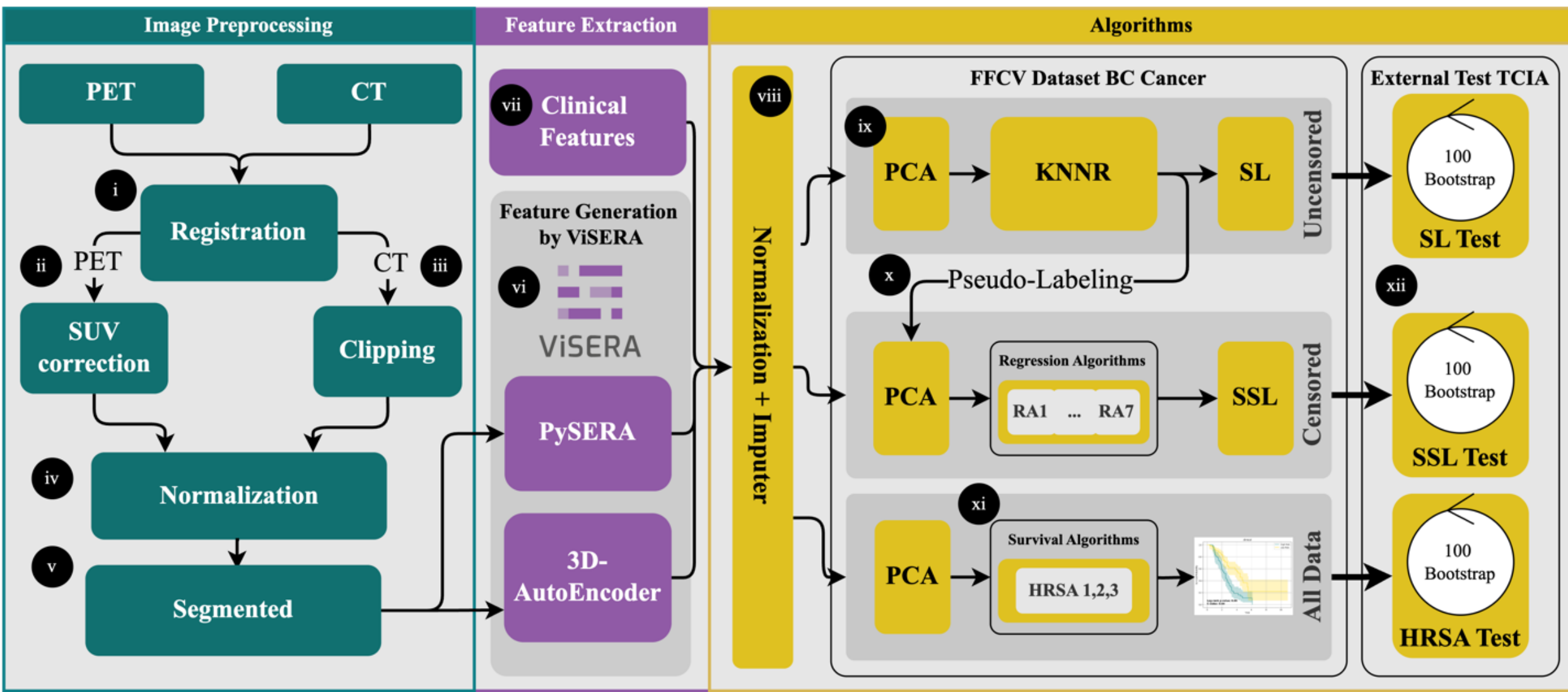

**Figure 1.** The study workflow consists of the following steps: (i) Positron Emission Tomography (PET) image registration on Computed Tomography (CT), (ii) Standardized Uptake Value (SUV) correction for PET images, (iii) clipping of CT images, (iv) normalization of PET and CT images, (v) image segmentation, (vi) Handcrafted Radiomics Feature (HRF) extraction using ViSERA and Deep Radiomics Feature (DRF) extraction using ViSERA 3D AutoEncoder, (vii) incorporating clinical data, (viii) combining clinical data with HRFs and DRFs, followed by normalization and imputation (ix) applying Supervised Learning (SL) in five-fold cross-validation (FFCV), (x) pseudo-labeling censored data using K-Nearest Neighbor Regression (KNNR) then applying Principle Component Analysis (PCA) and 7 regressors in FFCV in Semi-Supervised Learning (SSL) approach, (xi) conducting survival analysis with PCA and 3 Hazard Ratio Survival Analysis (HRSA), and (xii) validating all supervised, semi-supervised and survival risk assessment models with 100 bootstraps from external test data.

## 3. Results

### 3.1. PET imaging Regression

As shown in Table 2, the best performance for PET imaging was achieved using PET_HRF, where the SL strategy with PCA and the DTC algorithm resulted in an MAE of 2.11±0.46 years (outcome range: [0.1 - 6.6] years), while the SSL strategy improved to 1.55±0.25 years. The SSL approach showed a 26.54% improvement over SL ($p<0.05$, paired t-test). For the DRF model, combining PET_DRF and CF with MLP and PCA in the SSL strategy achieved an MAE of 2.07±0.49 years, compared to 2.52±0.47 years in the SL strategy. The SSL method demonstrated a 17.86% performance increase over SL ($p<0.05$, paired t-test). Additionally, using CF alone with MLP and PCA, the SSL strategy achieved an MAE of 1.76±0.49 years, while the SL strategy yielded an MAE of 2.18±0.47 years. The SSL strategy outperformed SL by 19.27% ($p<0.05$, paired t-test). Figure 2 presents the results using a violin plot, clearly illustrating the difference between the two approaches. Comprehensive regression results are provided in Supplemental Table S2. In summary, the SSL strategy significantly outperformed the SL strategy for both DRF and HRF models, as well as CF in PET imaging.

**Table 2**. The best performances provided by semi-supervised (SSL) vs. supervised learning (SL) approaches; DRF_PET: Deep Radiomics Features extracted from PET, HRF_PET: Handcrafted Radiomics Features Extracted From PET, CF: Clinical Features, MAE: Mean Absolute Error, MSE: Mean Squared Error, FFCV: Five-Fold Cross-Validation, DTR: Decision Tree Regressor, MLPR: Multi-Layer Perceptron Regressor, KNNR: K Nearest Neighbor Regressor, ABR: AdaBoost Regressor.

| Regressor | Feature Type | SL | | SSL | | p-value | %Gain |
|---|---|---|---|---|---|---|---|
| | | FFCV MAE | Test MAE | FFCV MAE | Test MAE | | |
| DTR | HRF_PET | 1.00 | 2.11 | 0.75 | 1.55 | <0.05 | 26.5% |
| MLPR | CF | 1.02 | 2.18 | 0.62 | 1.76 | <0.05 | 19.3% |
| MLPR | DRF_PET+ CF | 0.95 | 2.52 | 0.66 | 2.07 | <0.05 | 17.9% |
| DTR | HRF_PET+ CF | 1.01 | 2.20 | 0.72 | 1.84 | <0.05 | 16.4% |
| KNNR | DRF_PET | 0.94 | 2.47 | 0.59 | 2.17 | <0.05 | 12.2% |
| ABR | DRF_PET+ HRF_PET | 1.00 | 2.32 | 0.62 | 2.10 | <0.05 | 9.48% |
| MLPR | DRF_PET+ CF + HRF_PET | 1.03 | 2.37 | 0.76 | 2.19 | <0.05 | 7.59% |

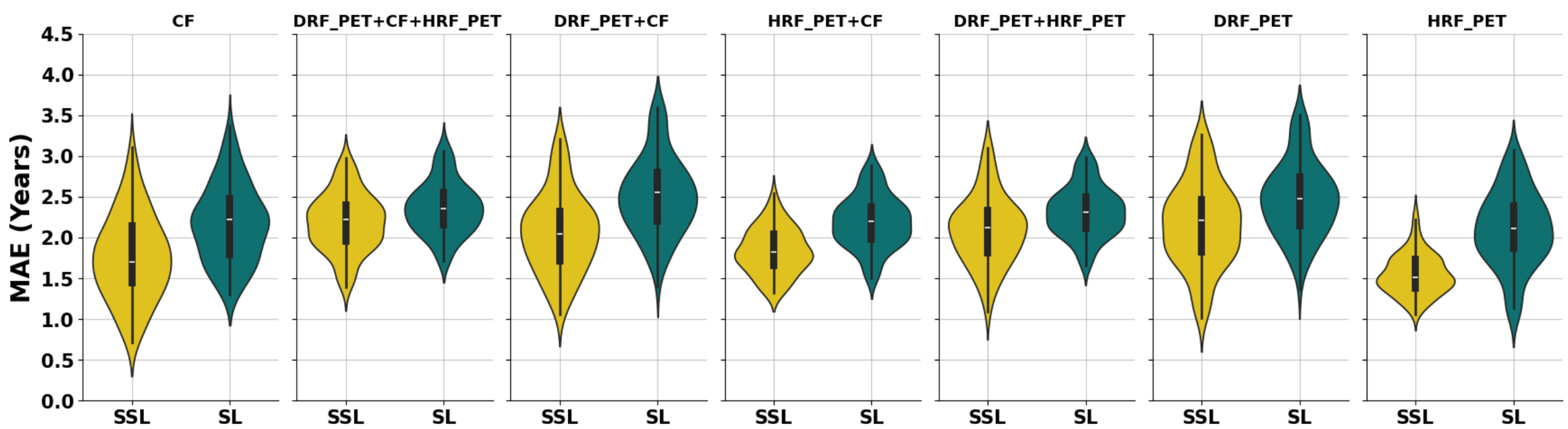

**Figure 2**. Violin plot for PET and clinical data demonstrating SSL vs SL; DRF_PET: Deep Radiomics Features extracted from PET, HRF_PET: Handcrafted Radiomics Features Extracted From PET, CF: Clinical Features, SL: Supervised Learning, SSL: Semi-Supervised Learning, MAE: Mean Absolute Error

### 3.2. CT imaging Regression

As shown in Table 3, for datasets with imaging features extracted from CT images, the optimal performance was achieved with the SSL strategy, yielding an MAE of 2±0.36 years using CT_DRF and CF combined with PCA and MLP. In comparison, the SL strategy employing the same imaging features resulted in an MAE of 2.12±0.49 years, showing a 6.28% improvement with SSL, which was slightly significant (p=0.053, paired t-test). For the HRF in CT, whether used alone or with clinical features, the MAE was 2.14 and 2.25, respectively, which were worse than the SL results of 2.04 and 2.05. Figure 3 visualizes the results through a violin plot. Comprehensive regression performances are shown in Supplemental Table S2. Our findings indicate that for CT imaging, SSL offers minimal improvement for the DRF framework and no improvement for the HRF framework.

**Table 3**. CT and clinical data SSL vs SL results; DRF_CT: Deep Radiomics Features extracted from CT, HRF_CT: Handcrafted Radiomics Features Extracted From CT, CF: Clinical Features, SL: Supervised Learning, SSL: Semi-Supervised Learning, MAE: Mean Absolute Error, MSE: Mean Squared Error, FFCV: Five-Fold Cross-Validation, MLPR: Multi-Layer Perceptron Regressor, KNNR: K Nearest Neighbor Regressor

| Regressor | Feature Type | SL | | SSL | | p-value | %Gain |
|---|---|---|---|---|---|---|---|
| | | FFCV MAE | Test MAE | FFCV MAE | Test MAE | | |
| MLPR | CF | 1.02 | 2.18 | 0.62 | 1.76 | <0.05 | 19.27% |
| KNNR | DRF_CT+HRF_CT | 1.05 | 2.26 | 0.61 | 2.08 | 0.012 | 7.96% |
| KNNR | DRF_CT+CF+ HRF_CT | 1.05 | 2.23 | 0.63 | 2.09 | 0.061 | 6.28% |
| MLPR | DRF_CT+CF | 1.00 | 2.12 | 0.73 | 2.00 | 0.053 | 5.66% |
| MLPR | DRF_CT | 0.99 | 2.14 | 0.76 | 2.12 | 0.767 | 0.93% |
| KNNR | HRF_CT | 1.04 | 2.04 | 0.57 | 2.14 | 0.142 | -4.90% |
| KNNR | HRF_CT+CF | 1.03 | 2.05 | 0.56 | 2.25 | 0.002 | -9.76% |

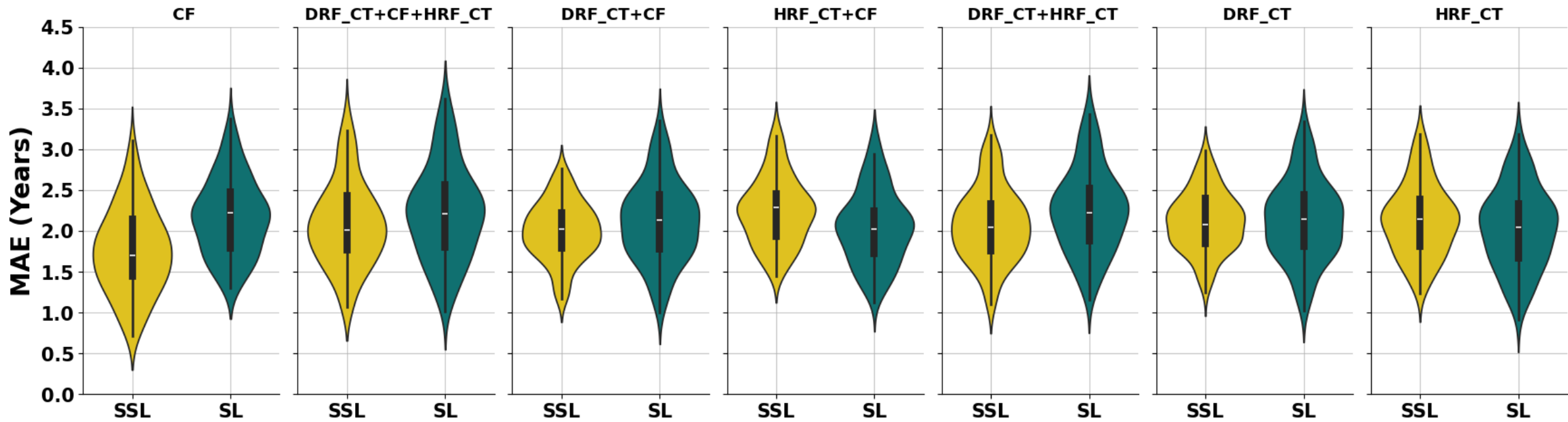

**Figure 3**. Violin plot for CT and clinical data demonstrating SSL vs SL; DRF_CT: Deep Radiomics Features extracted from CT, HRF_CT: Handcrafted Radiomics Features Extracted From CT, CF: Clinical Features, SL: Supervised Learning, SSL: Semi-Supervised Learning, MAE: Mean Absolute Error

### 3.3. Results of Hazard Ratio Survival Analysis (HRSA)

In our survival prediction models unlike the regression approach, as shown in Table 4, CT-based models demonstrated superior predictive accuracy in external tests. Specifically, the CT_HRF model yielded a c-index of 0.656 and a statistically significant p-value of 0.019 using the PCA+CWGBSA approach. For datasets incorporating PET, the PET_HRF+CF model achieved the best results with the PCA+FSSVM approach, resulting in an external c-index of 0.59 and a p-value of 0.06, which approached significance. In the clinical dataset, the survival model achieved a c-index of 0.63. However, the log-rank test did not reveal a significant difference between high-risk and low-risk groups, with a p-value of 0.44. The Kaplan-Meier survival curves corresponding to the highest-performing models are presented in Figure 4, illustrating the survival probability stratified by risk groups. Additional details on the survival analysis, including methodological specifics, can be found in Supplemental Table S3. Kaplan-Meier survival curves corresponding to all results are provided in Supplemental Figures S2–S4.

**Table 4.** Survival analysis results; HRSA: Hazard Ratio Survival Analysis, PCA: Principal Component Analysis, DRF_CT: Deep Radiomics Features extracted from CT, HRF_CT: Handcrafted Radiomics Features Extracted From CT, DRF_PET: Deep Radiomics Features extracted from PET, HRF_PET: Handcrafted Radiomics Features Extracted From PET, CF: Clinical Features, FFCV: Five-Fold Cross-Validation, STD: Standard Deviation, FSSVM: Fast Survival Support Vector Machine, CWGBSA: Component Wise Gradient Boosting Survival Analysis, RSF: Random Survival Forest

| Dataset | PCA+ HRSA | FFCV | | External Test | |
|---|---|---|---|---|---|
| | | c-index ± STD | p-value | c-index | p-value |
| HRF_CT | CWGBSA | 0.63±0.08 | 0.002 | 0.656±0.02 | 0.019 |
| HRF_CT + CF | CWGBSA | 0.63±0.08 | 0.002 | 0.653±0.03 | 0.118 |
| CF | FSSVM | 0.64±0.02 | 0.001 | 0.633±0.02 | 0.449 |
| DRF_CT + HRF_CT | RSF | 0.66±0.05 | <0.05 | 0.632±0.04 | 0.585 |
| HRF_PET + CF | FSSVM | 0.66±0.07 | <0.05 | 0.598±0.02 | 0.060 |
| DRF_CT + HRF_CT +CF | CWGBSA | 0.66±0.06 | 0.002 | 0.598±0.05 | 0.632 |
| DRF_PET + HRF_PET | RSF | 0.63±0.06 | 0.02 | 0.587±0.03 | 0.207 |
| HRF_PET | RSF | 0.65±0.06 | 0.002 | 0.573±0.04 | 0.628 |
| HRF_PET + CF | RSF | 0.65±0.06 | <0.05 | 0.569±0.02 | 0.107 |
| DRF_PET + HRF_PET +CF | RSF | 0.62±0.09 | 0.173 | 0.56±0.05 | 0.004 |
| DRF_CT + CF | RSF | 0.61±0.02 | 0.025 | 0.556±0.06 | 0.027 |
| DRF_CT | RSF | 0.61±0.05 | 0.014 | 0.553±0.03 | 0.075 |
| DRF_PET + CF | RSF | 0.58±0.05 | 0.021 | 0.513±0.02 | 0.371 |

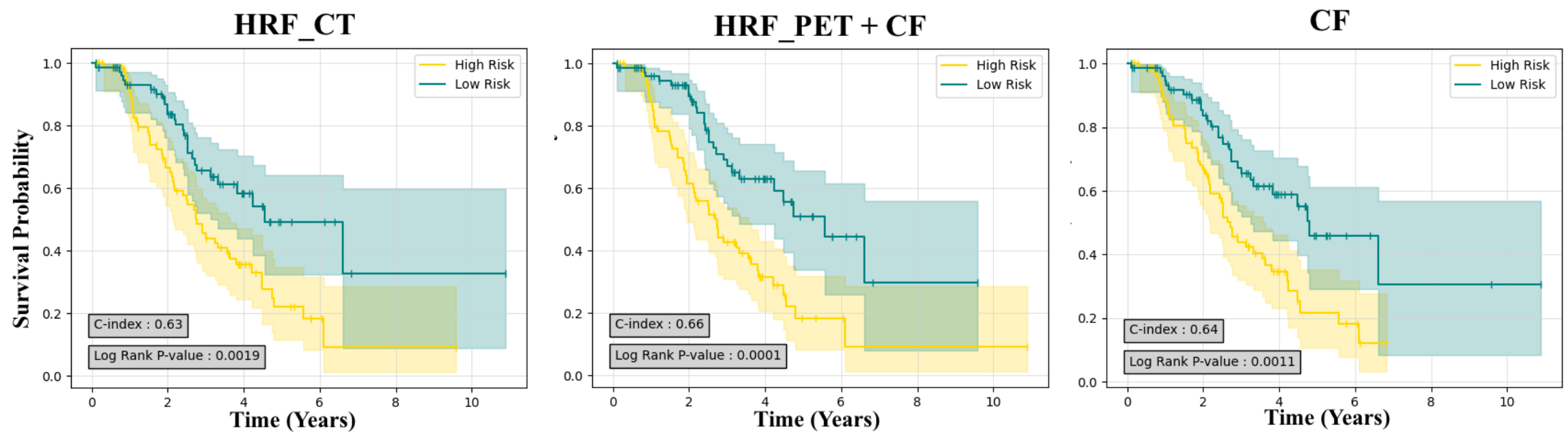

**Figure 4.** Kaplan-Meier survival curves generated for best performance in PET and CT and clinical data (CF); HRF_CT: Handcrafted Radiomics Features Extracted From CT, HRF_PET: Handcrafted Radiomics Features Extracted From PET

## 4. Discussion

The SSL strategy is gaining significant attention in cancer research, particularly for lung cancer prognosis, due to challenges in obtaining labeled datasets [55]. While many studies have applied SSL approaches to tasks such as PD-L1 scoring [55] or nodule detection [56, 57], few have focused on using SSL for survival prediction. Chai et al. employed an SSL framework (Cox-AFT model) that combined CoxPH and AFT models to predict survival time using genomics data across various cancers, including lung cancer, achieving a 15% reduction in MSE, which highlights the benefits of SSL for genomics data [58]. Hermoza et al. applied pseudo-labeling to predict survival time from pathology and X-ray images, achieving a 21% improvement in MAE despite varying levels of censored data [59]. Similarly, Nateghi and Vens demonstrated improved predictive performance in survival tasks by including unlabeled data with transcriptomics and clinical data [60]. In a recent study, we utilized pseudo-labeled data from head and neck cancer to enhance the prediction accuracy of two-year lung cancer survival, achieving a significant improvement in predictive performance [22]. Our findings align with these studies, showing that SSL significantly outperformed SL, particularly in PET-based models. Specifically, applying SSL to PET_HRF resulted in a 26.54% improvement in MAE, while the CF model showed a 19.27% improvement in MAE. By incorporating both censored and uncensored data, our approach reduces bias and enhances the model's generalizability across diverse patient populations, ultimately improving the accuracy and clinical utility of survival predictions.

In the context of survival prediction, this study demonstrates that CT-based imaging models exhibit superior predictive accuracy compared to PET imaging and clinical data alone. Among the CT-based models, the highest performance was achieved by the CT_HRF model, with a C-index of 0.656±0.02 and a statistically significant p-value. For PET-based models, the best performance was observed with the PET_HRF+CF model, which attained a C-index of 0.598±0.02 and a p-value of 0.06. PET imaging may be more effective in predicting exact survival times using censor-aware semi-supervised approaches, potentially due to its ability to capture subtle metabolic variations that might align closely with time-to-event data [61]. Conversely, CT could perform better in survival prediction tasks, as its detailed structural features might provide patterns that advanced machine learning techniques are better suited to leverage for stratifying risk and predicting broader survival outcomes [62].

The advancement of personalized medicine in cancer management has underscored the unique advantages of PET and CT imaging. Several studies [63, 64, 65, 66] have demonstrated that convolutional neural networks trained on pre-treatment PET/CT images can effectively predict lung malignancy progression and OS. Additionally, radiomic signatures derived from PET/CT images have been shown to predict disease-free survival in NSCLC patients [67]. In a study by Kirienko et al., disease-free survival prediction using radiomic and CF revealed that CT performed slightly better than PET [68]. Other research [69] identified variables from CT and PET/CT as strong survival predictors, with Liu et al. [70] suggesting that CT radiomics may outperform PET in certain scenarios. In contrast, our analysis found that PET imaging demonstrated better baseline performance in the SL strategy and achieved greater performance improvements when using the SSL strategy compared to CT. This discrepancy may be partially attributed to the small sample size and variability in the quality of our dataset, which could influence the relative performance of PET and CT imaging.

Our study contributes to the field by integrating diverse feature types and employing both deep learning and traditional feature extraction methods. Recent research [71, 72, 36] highlights the potential of DRFs and tensor radiomics to enhance outcome prediction. Notably, Beraghetto et al. demonstrated that HRFs outperformed DRFs in predicting two-year survival, which aligns with our findings for PET imaging. However, in our study, DRFs achieved better performance for CT imaging. While DRFs are advantageous due to their simplicity, as they are automatically learned from imaging data using deep learning models, eliminating the need for segmentation and reducing human biases inherent in HRFs, they often involve managing a larger number of features compared to HRFs. For small sample sizes, as in our study, this higher feature complexity can disadvantage DRFs and favor the more concise and interpretable HRFs [73, 33]. This underscores the importance of considering dataset size and feature complexity when selecting predictive approaches for survival analysis.

Our study has several limitations that offer opportunities for future research. First, the datasets were limited to specific cancer types (lung cancer) and imaging modalities (CT and PET), restricting generalizability. Validation with diverse datasets, including head and neck, liver, and prostate cancers, is needed to enhance robustness. The SSL approach relied on pseudo-labeling, which may introduce noise, emphasizing the importance of high-quality pseudo-labels. While DRFs effectively capture complex patterns, they lack the interpretability of HRFs. The study's small sample size necessitates validation with larger, independent datasets. Feature selection relied on PCA to prevent overfitting, but advanced methods could further optimize performance. In survival prediction, SSL was not feasible due to its dependency on the last follow-up date, limiting broader applicability. Manual tumor segmentation required expert input, highlighting the potential for automated deep learning-based segmentation. Additionally, the study employed a limited range of RAs and HRSAs; exploring more algorithms could improve predictive accuracy. Future work should focus on diverse datasets with complete follow-up information, additional imaging modalities, and advanced techniques, such as explainable machine learning models, to address these limitations and further advance the field.

## 5. Conclusion

Our findings demonstrate that the implementation of censor-aware semi-supervised survival time prediction significantly enhances model performance compared to supervised learning. This enhancement is particularly notable when applied to radiomic features extracted from PET imaging, with the PET_HRF model achieving the best performance, showing a 26.5% improvement in MAE. This approach effectively addresses the challenges posed by right-censored data and mitigates potential biases inherent in traditional survival prediction methods. By integrating this methodology into survival analysis frameworks, we establish a robust and generalizable model capable of delivering accurate predictions even with limited survival data. This capability is particularly valuable in clinical settings where data availability is often constrained. Specifically, our findings can assist clinicians in identifying high-risk patients, tailoring treatment strategies, and optimizing patient outcomes, particularly in oncology settings where personalized approaches are critical for improving survival and quality of care.

**Data and Code Availability.** All code (including prediction and dimension reduction algorithms) is publicly shared at: *https://github.com/MohammadRSalmanpour/Several-Novel-Machine-Learning-Strategies-for-Lung-Cancer-Survival-Outcome-Prediction*

**Acknowledgements.** This study was supported by Virtual Collaboration (VirCollab, *www.vircollab.com*) Group as well as Technological Virtual Collaboration Corporation (TECVICO CORP.), Vancouver, Canada. This study was supported by the support of the Natural Sciences and Engineering Research Council of Canada (NSERC) Discovery Horizons Grant [DH-2025-00119] and UBC Department of Radiology 2023 AI Fund.

**Ethics Statements.** This study was approved by the University of British Columbia - BC Cancer Research Ethics Board. Study ID: H19-02805 - "Utilizing Lung Cancer Databases to Validate Outcome Prediction Models, A Pilot Study", Date: 20 April 2023. The requirement for individual informed consent was waived due to the retrospective nature of the study and the use of anonymized data. Other public datasets were provided from TCIA, https://www.cancerimagingarchive.net/collection/tcga-lusc/.

**Conflict of Interest.** Mohammad R. Salmanpour is affiliated with TECVICO Corp. The remaining authors declare no relevant conflicts of interest.